# Barometric Altimeter Assisted SINS/DR Combined Land Vehicle Gravity Anomaly Method

Kefan Zhang
Rocket Force University of Engineering
The Key Laboratory of Communication Engineering
Xi'an, China
kefanzhang2024RFUE@outlook.com

Zhili Zhang
Rocket Force University of Engineering
The Key Laboratory of Communication Engineering
Xi'an, China
zhangzl_xiht@163.com

Junyang Zhao*
Rocket Force University of Engineering
The Key Laboratory of Communication Engineering
Xi'an, China
* Corresponding author: zhaojy802@sina.com

Shenhua Lv
Rocket Force University of Engineering
The Key Laboratory of Communication Engineering
Xi'an, China
1368611795@qq.com

*Abstract*—Traditional land vehicle gravity measurement heavily rely on high-precision satellite navigation positioning information. However, the operational range of satellite navigation is limited, and it cannot maintain the required level of accuracy in special environments. To address this issue, we propose a novel land vehicle gravity anomaly measurement method based on altimeter-assisted strapdown inertial navigation system (SINS)/dead reckoning (DR) integration. Gravimetric measurement trials demonstrate that after low-pass filtering, the new method achieves a fit accuracy of 2.005 mGal, comparable to that of the traditional SINS/global navigation satellite system (GNSS) integration method. Compared with the SINS/DR integration method, the proposed method improves accuracy by approximately 11%.

## I. INTRODUCTION

The primary methods for terrestrial disturbance gravity densification include ground-based single-station static gravimetry, airborne gravimetry, and land vehicle gravimetry[1]. Among these, ground-based single-station static gravimetry offers the highest precision. However, this method is limited by geographical environmental constraints and has a very low measurement efficiency. For an equivalent survey area, the time cost of static gravimetry is 4 to 6 times that of dynamic gravimetry[2]. Airborne gravimetry is the primary means for gravity surveys in areas difficult to access from the ground, providing rapid and efficient measurement of large-scale perturbation gravity data. However, this method can only measure perturbation gravity along flight lines, often requiring downward continuation for complete coverage[3]. In contrast, land vehicle gravimetry is conducted directly at the Earth's surface, avoiding the accuracy loss associated with downward continuation. Moreover, because vehicle-based gravimetry is closer to the gravity source, it can yield high signal-to-noise ratio and high-resolution gravity information, even at lower speeds. Currently, land vehicle gravimetry technology has become the predominant method for terrestrial gravity surveys[4].

Global Navigation Satellite Systems (GNSS) have long played a critical role in land vehicle gravimetry[5]. Firstly, GNSS position and velocity information are essential for compensating errors in strapdown inertial navigation systems (SINS), which enhances the precision of specific force measurements. Secondly, the carrier's acceleration and various corrections rely on GNSS positioning data. However, during terrestrial gravimetric surveys, vehicles often encounter complex GNSS observation environments such as valleys, tunnels, and overpasses. This leads to degraded GNSS observation quality, which directly affects the accuracy of gravimetric measurements[6].To reduce the dependence of land vehicle gravimetry on GNSS positioning, a SINS/velocity sensor (VEL) integrated system for land vehicle gravimetry was proposed in [7]. However, this method does not provide direct height measurements, which limits the accuracy of the gravimetric data. Subsequently, Literature [8] introduced a land vehicle gravimetry approach based on the integration of SINS with a laser Doppler velocimeter (LDV). The use of LDV height information improved the accuracy of gravimetric measurements. Nonetheless, the velocity measurement accuracy of LDVs is influenced by ambient light intensity and vehicle speed[9], posing significant challenges to the resolution and reliability of this method.

Addressing the limitations of previous methods, this paper proposes a barometric altimeter (BALT)-assisted strapdown inertial navigation system (SINS)/dead reckoning (DR) combination for vehicular dynamic gravimetry, hereafter referred to as the SINS/DR+BALT method. This approach leverages the SINS/DR integrated navigation to compensate for the reliance on GNSS positioning or optical velocimeter velocity information in traditional gravimetric surveys. By incorporating high-precision altitude data from the BALT, the method achieves combined altitude measurements, thereby significantly enhancing the accuracy of gravimetric measurements.

## II. MATHEMATICAL MODEL

The vehicular gravimetry discussed herein refers to scalar gravimetry, with the physical quantity measured being the gravity anomaly. In the local navigation frame, the mathematical model for the gravity anomaly is given by[10]:

$$\delta g_U = \dot{v}_U - f_U + \delta a_E - \gamma \quad (1)$$

where, $\delta g_U$ is the gravity anomaly; $\dot{v}_U$ is the upward component of vector acceleration; $f_U$ is the upward component of specific force; $\delta a_E$ and $\gamma$ are Eotvos corrections and normal gravity, respectively:

$$\delta a_E = 2\omega \cdot v_E \cdot \cos L + \frac{v_E^2}{R_N + h} + \frac{v_N^2}{R_M + h},$$

$$\gamma = \frac{R_e g_e \cos^2 L + R_p g_p \sin^2 L}{\sqrt{R_e^2 \cos^2 L + R_p^2 \sin^2 L}}.$$

$\omega$ is the angular speed of the Earth's rotation; $v_E$ and $v_N$ are the carrier's east velocity and north velocity respectively; $L$ is the latitude; $R_N$ and $R_M$ are the equatorial major semi-axis and the polar minor semi-axis of the rotating ellipsoid model; $h$ is the height; $R_e$ and $R_p$ are the equatorial major semi-axis and the polar minor semi-axis of the rotating ellipsoid model; $g_e$ and $g_p$ are the magnitudes of equatorial gravity and pole gravity.

## III. DATA PROCESSING FLOW

Dynamic gravimetry comprises two main components: combined navigation and gravity extraction. The latter includes obtaining the raw gravity anomaly using (1) and applying low-pass filtering. The specific steps are summarized as follows:

*a)* SINS performs the initial alignment and forms the dead reckoning system with the odometer.

*b)* The inertial navigation is solved by SINS, and the position information of DR Is used as the observation quantity for SINS/DR Kalman filter integrated navigation, and the combined specific force, attitude, velocity and position information is output.

In SINS/DR Integrated navigation, the state vector $\mathbf{X}$ has a total of 21 dimensions, including attitude error, speed error, position error, inertial device error sum, position error of dead reckoning, odimeter scale coefficient error and installation error Angle of SINS.

The residual difference between the SINS output position and the dead reckoning position is taken as measurement, and the measurement equation of the system is as follows:

$$\mathbf{Z} = \mathbf{HX} + \mathbf{V}$$
$$\mathbf{Z} = \begin{bmatrix} \mathbf{P}_{SINS}^n - \mathbf{P}_{DR}^n \end{bmatrix} \quad (2)$$
$$\mathbf{H} = \begin{bmatrix} \mathbf{0}_{3\times 6} & \mathbf{I}_{3\times 3} & -\mathbf{I}_{3\times 3} & \mathbf{0}_{3\times 9} \end{bmatrix}$$

*c)* The attitude, latitude and velocity information of integrated navigation are used to calculate the specific force, normal gravity and Eotvos correction. The combined height is obtained by using the combined altimetry scheme, and the acceleration of carrier motion is obtained by quadratic difference.

The combined altimeter scheme uses the weighted sum of altimeter height and dead reckoning height as the combined height. The weight coefficient is determined by the measurement variance of DR And BALT:

$$h_M = \frac{\sigma_D^2}{\sigma_D^2 + \sigma_B^2} h_{BALT} + \frac{\sigma_B^2}{\sigma_D^2 + \sigma_B^2} h_{DR} \quad (3)$$

where, $h_M$ is the combined height; $\sigma_D^2$ and $\sigma_B^2$ are the measurement variance of DR and barometric altimeter; $h_{BALT}$ and $h_{DR}$ are barometric altimeter height and dead reckoning height.

*d)* The original gravity anomaly is obtained by substituting each correction and carrier acceleration into (1).

*e)* A 100s FIR low-pass filter is used to remove the high-frequency noise in the original gravity anomaly[11].

*f)* The gravity results calculated by EGM2190 model were used as reference values to evaluate the external coincidence accuracy. The accuracy evaluation formula is:

$$\rho = \pm \sqrt{\frac{\sum_{i=1}^{N} \triangle g_i - \text{REF}_i}{N}}$$

where, $\rho$ is the off-line coincidence accuracy; $\triangle g_i$ is the gravity anomaly measured at sampling point $i$; $\text{REF}_i$ is the gravity anomaly at $i$ calculated for EGM2190 model; $N$ is the number of sampling points.

Figure 1 shows the flow diagram of the SINS/DR+BALT method.

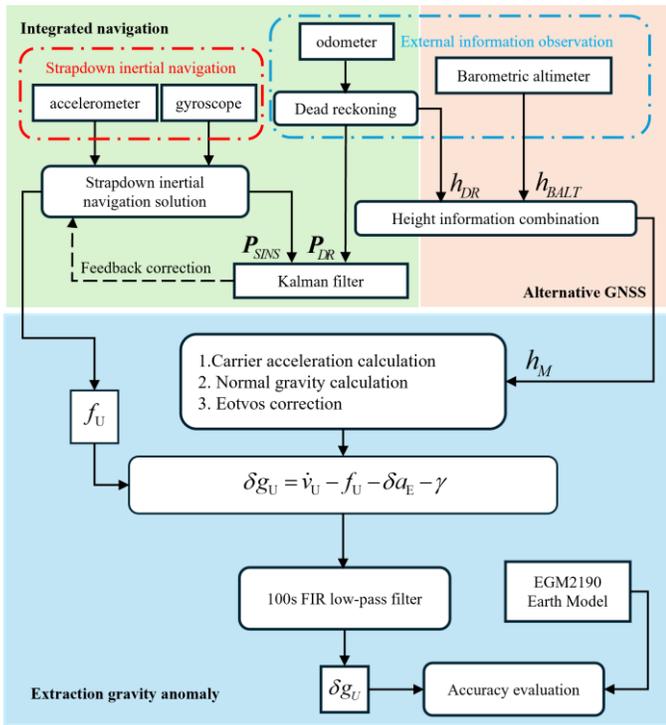

Figure 1 Data processing flow of SINS/DR+BALT method.

## IV. EXPERIMENTAL VERIFICATION AND ANALYSIS

In June 2024, the project team conducted a single-line gravity measurement experiment on a highway in Tianjin. In this experiment, Cheetah SUV was used as the carrier, and the core equipment included strapdown inertial navigation, odometer and altimeter (see Figure 2). Among them, the velocity measurement accuracy of the odometer is 0.001 m/s, and the measurement accuracy of the altimeter is better than 1 m.

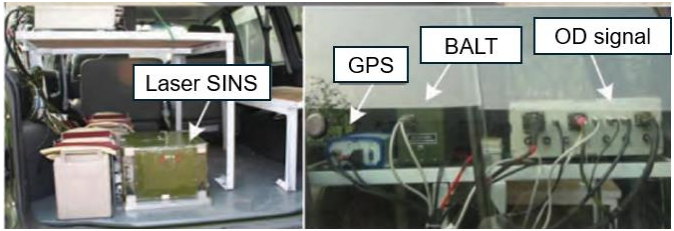

Figure 2 Vehicle gravity measurement system.

Figure 3 is the trajectory diagram of this vehicle gravity measurement test. The coordinates are relative latitude and longitude. The SINS/DR/BALT method is used as the experimental group, and the SINS/DR And SINS/GNSS methods are used as the control group.

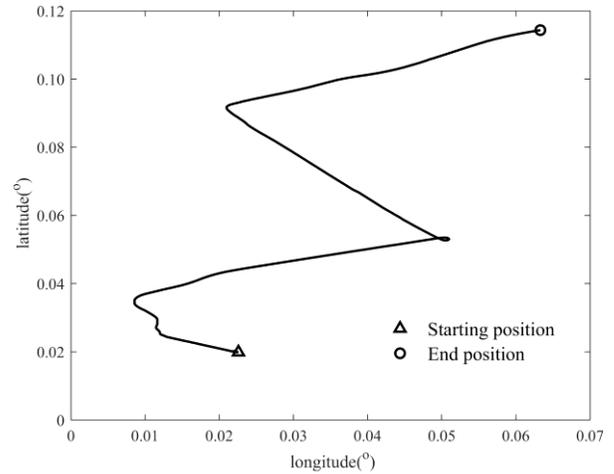

Figure 3 Test track.

Figure 4, Figure 5 and Figure 6 are gravity measurement results of SINS/GNSS method, SINS/DR Method and SINS/DR+BALT method respectively.

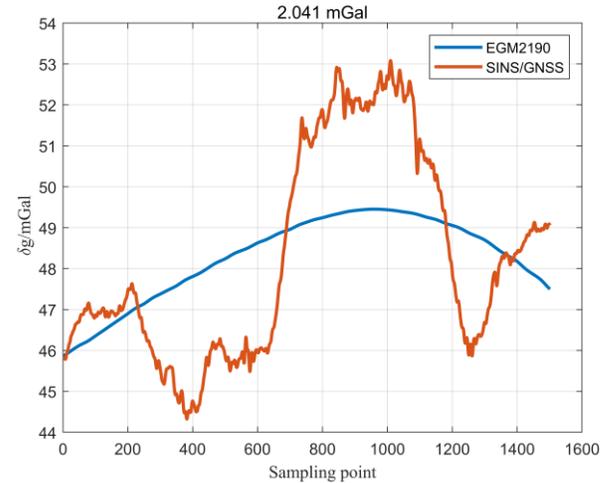

Figure 4 Results of SINS/GNSS gravity anomaly: After 100 s lowpass filtering, the outer matching accuracy is 2.041 mGal,

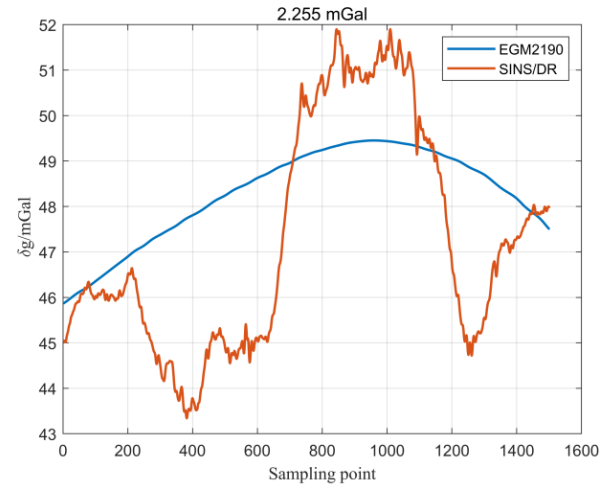

Figure 5 SINS/DR Gravity anomaly results: After 100 s lowpass filtering, the accuracy of external coincidence is 2.255mGal.

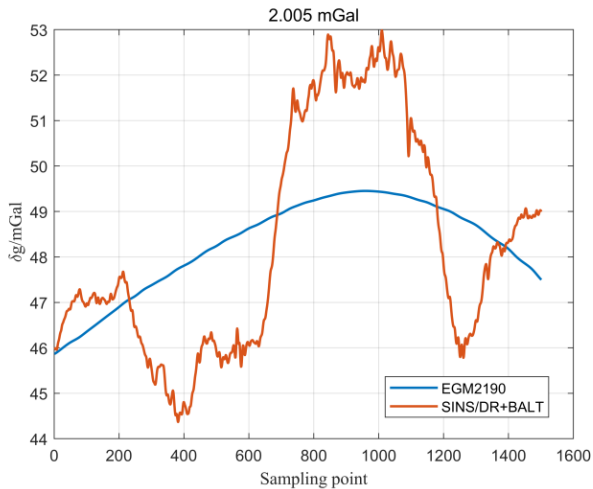

Figure 6 Results of SINS/DR+BALT gravity anomaly: After 100 s lowpass filtering, the external coincidence accuracy is 2.005 mGal.

As can be seen from Figure 4~6, the measurement accuracy of SINS/GNSS method in this experiment is 2.04mGal. The measurement accuracy of SINS/DR+BALT is 2.01 mGal. The external coincidence accuracy of the proposed method is comparable to that of SINS/GNSS method. In addition, compared with SINS/DR+BALT method, the external coincidence accuracy of SINS/DR Method is improved from 2.26 mGal to 2.01 mGal, and the accuracy is improved by about 11%.

## V. Conclusions

The strapdown SINS/DR+BALT vehicle gravity measurement method is studied in this paper. By analyzing the mathematical model of vehicle gravity measurement, the data processing flow of this method is given. The vehicle-mounted test proves that the proposed method can complete the dynamic gravity measurement test without GNSS positioning information. In addition, the DR/BALT combination altimetry scheme realizes the autonomous high precision measurement of height, which can further improve the accuracy of gravity measurement. In the next step, the effect of the velocity and latitude obtained by the combined navigation on the gravity measurement should be taken into account.

ACKNOWLEDGMENT

Thanks are given for the guiding comments from anonymous reviewers. This work was partially supported by the National Natural Science Foundation of China under Grant No. 62305393.

REFERENCES

[1] Li J., Wang S.H., Dai C.C., et al. (2018) Discussion on unsupported operational mapping support for missile. Aerospace Technology., 08:76-79.

[2] Rui H.Y. (2020) Research on Key Technologies for Strapdown Ground Vehicle Gravimetry. National Defense Industry Press, Xi'an.

[3] Gao S.J., Hao W.F., Li F. et al. (2018) Progress in Application of Airborne Gravity Measurements in Polar Regions. CHINESE JUORNAL OF POLAR REASEARCH., 30:97-113.

[4] Yu R.H., Cai S.K., Wu M.P., et al. (2015) A study of SINS/GNSS strapdown ground vehicle gravimetry test. Geophysical and Geochemical Exploration., 39:67-71.

[5] Yu R.H., Cai S.k., Wu M.p., et al. (2015) An SINS/GNSS Ground Vehicle Gravimetry Test Based on SGA-WZ02. Sensors., 15: 23477‑23495

[6] Yu R.H., Wu M.P., Cao J.L., et al. A new method of GNSS fault data detection for strapdown land vehicle gravimetry[A]. 2018 IEEE International Conference on Applied System Invention (ICASI)[C]. Chiba: IEEE, 2018: 299‑302.

[7] Yu R.H., Wu M.P., Zhang K.D., et al. (2017) A New Method for Land Vehicle Gravimetry Using SINS/VEL., Sensors, 17: 766.

[8] Wei G., Yang Z.K., Gao C.F. et al. (2023) Strapdown vehicle autonomous gravimetry method based on two-dimensional laser Doppler velocimeter. Infrared and Laser Engineering., 53:339-346.

[9] Zhang Z.L., Zhou Z.F. Chen H. (2021) Vehicle autonomous combination positioning and orientation technology. National Defense Industry Press, Xi'an.

[10] Xiong Z.M., Cao J.L., Wu M.P., et al. (2020) A Method for Underwater Dynamic Gravimetry Combining Inertial Navigation System, Doppler Velocity Log, and Depth Gauge. IEEE Geoscience and Remote Sensing Letters., 17(8): 1294‑1298.

[11] Cai T.J., Shao J.J., Hu X.L. (2023) Forward and Reverse FIR Filter Algorithm in Ocean Gravity Data Processing., 45:747-751.